\documentclass{vgtc}                          
\usepackage{float}
\usepackage{graphicx}

\graphicspath{{figures/}{pictures/}{images/}{./}} 

\usepackage{times}                     

\usepackage{booktabs}                  
\usepackage{lipsum}                    
\usepackage{mwe}                       

\usepackage{mathptmx}                  
\usepackage{amsmath}                   
\usepackage{amssymb}                   

\onlineid{0}

\vgtccategory{Research}

\vgtcinsertpkg

\title{Attention-based ROI Discovery in 3D Tissue Images}

\author{Hossein Fathollahian\thanks{e-mail:hfatho2@uic.edu}\\ %
        \scriptsize 
University of Illinois Chicago
\and Siyuan Zhao\\ %
     \scriptsize 
University of Illinois Chicago
 \and Nafiul Nipu\\ %
     \scriptsize 
University of Illinois Chicago
\and G. Elisabeta Marai\\ %
     \parbox{1.4in}{\scriptsize \centering University of Illinois Chicago}}
\teaser{
  \centering
  \includegraphics[width=\linewidth]{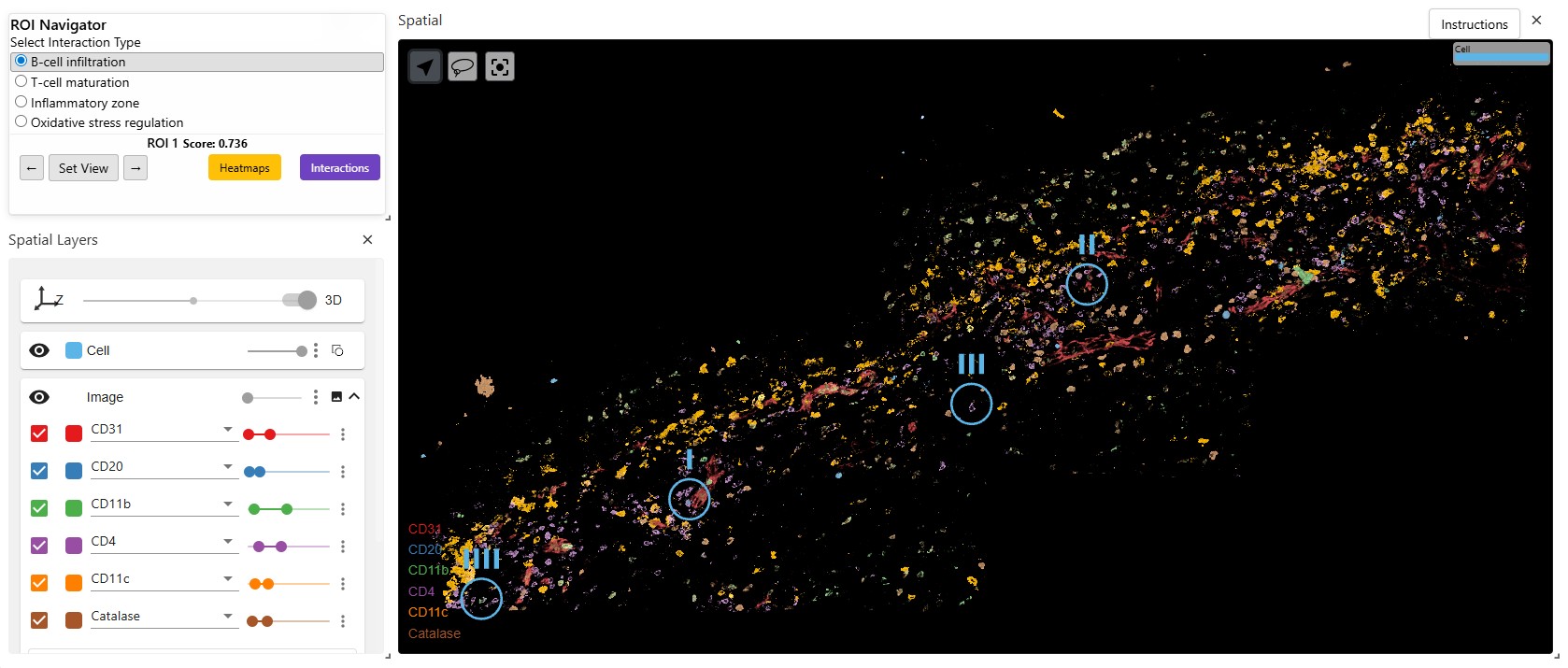}
  \caption{ Interactive ROI Identification through Biomarker Interaction with SSGAT: (left top) ROI navigator: selecting, showing, and displaying ROIs features; (left bottom) Marker control panel; (right) Multiplexed image showing tissue with marked regions of interest (ROIs).}
  \label{fig:teaser}
}
\abstract{
High-dimensional tissue imaging generates highly complex 3D data containing multiple biomarkers, making it challenging to identify biologically relevant regions without an expert user specifying manual labels for regions of interest. We introduce an approach to automatically identifying regions of interest (ROIs) in the 3D microscopy data. Our approach is based on a novel self-supervised multi-layer graph attention network (SSGAT), coupled with a React interactive interface wrapped around Vitessce. SSGAT uses an adversarial self-supervised learning objective to discover meaningful immune microenvironments across marker interactions. Our method reveals complex spatial bioreactions that can be visually assessed to assess their distribution across tissue.
    
} 
\begin{document}
\firstsection{Introduction}
\maketitle
Advancements in multiplexed imaging have opened up the possibility of high-resolution reconstruction of 3D tissues, allowing biomedical researchers to examine the location and extent of specific cell-cell interactions within a given volume~\cite{moore2021ome, hu2021spagcn}. The high data size and density, however, make it difficult to identify biologically relevant regions of interest (ROIs) without domain expert input and guidance~\cite{dong2022stagate, wu2022space,torkamani2015video}. 

To address this problem, we introduce a hybrid offline-online solution for automatically extracting ROIs and exploring these regions interactively (Fig.~\ref{fig:teaser}). We design a novel model, SSGAT, which is a Graph Attention Network (GAT) with Self-Supervised Learning that can identify interaction-rich ROIs in 3D images without requiring labeled data for training. We couple the results from SSGAT with a React wrapper around Vitessce to facilitate the interactive exploration of biomarker interactions in these ROIs.

\section{Related Work}
Most graph-based spatial analysis techniques leverage either manual labels or segmentation. SpaGCN~\cite{hu2021spagcn} spatially detects the domain in 2D transcriptomics using graph convolutional networks with an external segmentation that may result in degraded spatial accuracy. Other graph methods, such as STAGATE and SPACE-GM~\cite{dong2022stagate, wu2022space}, do not aim to detect ROI automatically, and are developed for 2D images only. In contrast, our work proposes an automated self-supervised learning mechanism that does not require manual labels and works on 3D data. Through a joint analysis of biomarker co-expression and spatial adjacency, SSGAT makes it possible to identify, on a scalable basis and in an unsupervised manner, biologically meaningful ROIs in immune microenvironments.
\begin{figure*}[tb]
  \centering
  \includegraphics[width=\linewidth]{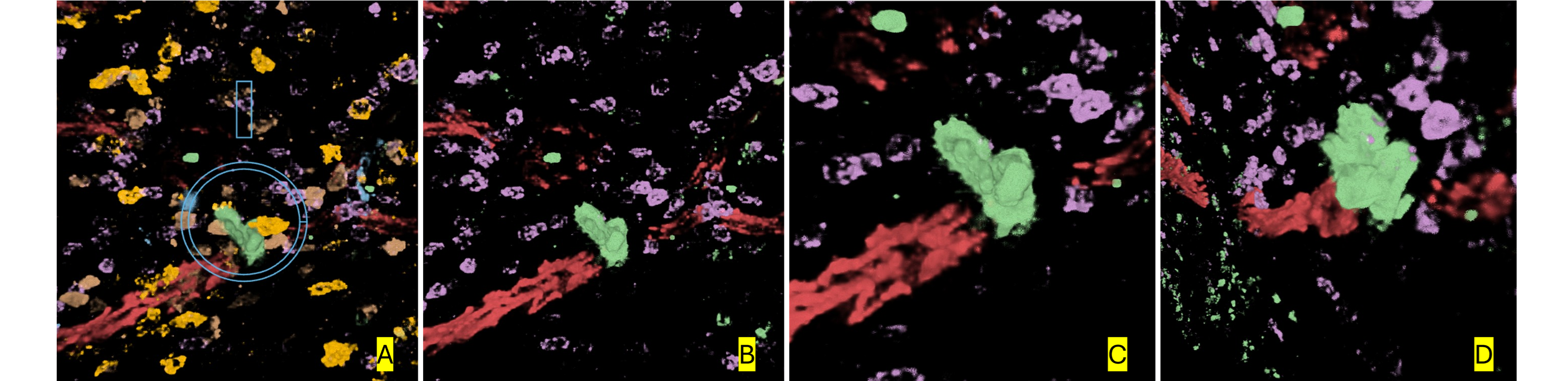}
  \caption{In-depth interactive analysis of a representative inflammatory ROI:
(A) Top-ranked inflammatory ROI identified by SSGAT, showing all biomarkers;
(B) Visualization of key inflammation-related biomarkers within the ROI: CD11b (Green), CD4 (Purple), and CD31 (Red); (C) Magnified and slightly tilted view of the ROI to highlight spatial co-localization and marker interactions;
(D) A rear view of the same ROI providing additional spatial and structural context within the tissue.}
  \label{fig:ROIs}
\end{figure*}
\section{Methodology}
\begin{figure}[H]
 \centering 
 \includegraphics[width=\columnwidth]{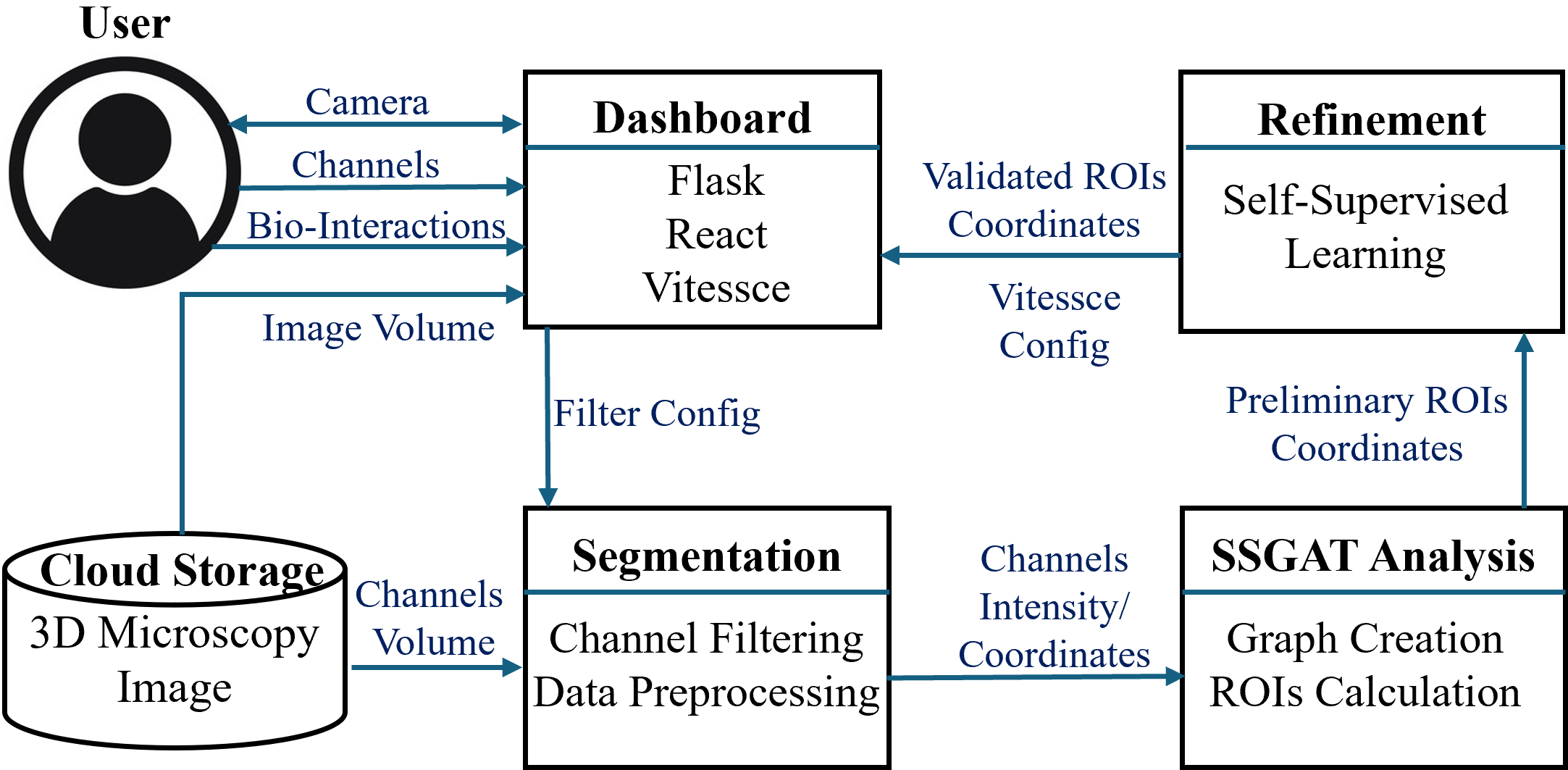}
 \caption{Architecture of our hybrid offline-online SSGAT solution.}

 \label{fig:architecture}
\end{figure}

Our solution (Fig.~\ref{fig:architecture}) comprises three offline modules: Segmentation, SSGAT Analysis, and Refinement,as well as an online module wrapper around Vitessce (described below). In the Segmentation module, user-set thresholds filter marker channels, coding significant voxels as 5D tuples (channel label, intensity, 3D coordinates) and condensing them into meaningful regions of interest for reliable graph construction. In the SSGAT Analysis module, described in more detail below, we construct spatial graphs from these tuples, represent intra- and inter-channel interactions as an SSGAT network, and use the network to identify and define potential regions of interest (ROIs). In the Refinement module, we use self-supervised learning to rank the list of detected potential ROIs bybiological relevance over the whole dataset. The offline pipeline is implemented in Python with numpy, torch, and torch\_geometric and processes a 3D Microscopy image in about some minutes. 
\noindent{\bf SSGAT Construction.} The SSGAT is constructed as follows: nodes are the biomarker intensities and weighed edges are the spatial relations between biomarkers (with the distance as the weight), which are connected to represent multi-layer GAT relationships. This joint intensity and distance analysis of nodes and edges captures dense regions with high intensities. We construct a two-layer Graph Attention Network: the first layer has four heads focused on proximity, similarity, complementarity, and spatial configuration, and the second layer has heads equal to the number of user-defined interactions. The module produces a list of ROIs (as voxel collections) ranked by the total SSGAT score of each ROI. SSGAT is the first attempt, to our knowledge, to introduce graph-based reasoning into 3D space and to implement concepts of contrastive learas well as local and global attention, allowing for allow the automated identification of immunity interplay regions. 

\noindent{\bf{Front end.}} We combine SSGAT with the Vitessce framework~\cite{keller2025vitessce} and OME-Zarr~\cite{moore2021ome}. The React user interface allows exploring ROIs at various zoom levels. The users select marker channels and specify desired biological interactions to direct the graph construction and analysis. ROIs are shown as overlay labels, and the zoom, rotation, and channel settings are configurable. Individual ROIs can be selected through the interface menu, and the camera point of view is automatically set to the ROIs centroid, offset by -1 on the Z axis. Single-channel and aggregated heatmaps of each defined interaction can be displayed as details on demand. 

\section{Results}
Figure~\ref{fig:ROIs} illustrates an example investigation of an automatically detected inflammatory ROI. Two additional examples are included in the supplemental materials, showing an automatically detected B-cell ROI, respectively an automatically detected T-cell ROI. For each supplemental example we show the context of the image, a selected channels only view, zoomed-in and rotated views highlighting the spatial interaction among biomarkers, and a close-up view with intensity heatmaps illustrating the expression patterns and interactions of the selected channels. A light version of our solution is available online: \url{https://hosseinfatho.github.io/SSGAT/}, showing precomputed ROIs, and where the channel and interaction heatmaps are disabled due to the computational cost.

\section{Conclusion}
We introduced a solution for automatically identifying ROIs based on SSGAT, a new model that can be used to calculate different biologically important ROIs in 3D multiplexed images of tissues.  Our solution shows an overview of all the ROIs, and supports individual ROI exploration through the camera position, zooming parameters, and the value range of the marker channels. In future work, we would like to couple the offline and online components, enhance the sensitivity to rare cellular patterns, increase generalizability by using weak supervision, facilitate automatic detection of molecular interactions based on the selection of channels, and make our approach compatible with complementary imaging modalities.
\acknowledgments{We thank the U.S. NIH (R01CA258827, UG3TR004501), NSF (CNS-2320261), and the UIC Institute for Health Data Science Research.}

\bibliographystyle{abbrv-doi}

\bibliography{template}
\end{document}